\documentclass[12pt]{article}
\usepackage[left=2.5cm,top=2.50cm,right=2.5cm,bottom=2.50cm]{geometry}
\usepackage{mathrsfs}
\usepackage{amsmath,amssymb,latexsym,color,cancel,graphicx,bbm,colortbl}
\usepackage[english]{babel}
\usepackage[latin1]{inputenc}
\usepackage{ragged2e}
\begin{document}
\date{}

\title{A generalized Jaynes-Cummings model: The relativistic parametric amplifier and a single trapped ion}
\author{D. Ojeda-Guill\'en$^{a}$,\footnote{{\it E-mail address:} dojedag@ipn.mx}\\ R. D. Mota$^{b}$ and V. D. Granados$^{c}$} \maketitle

\begin{minipage}{0.9\textwidth}
\small $^{a}$ Escuela Superior de C\'omputo, Instituto Polit\'ecnico Nacional,
Av. Juan de Dios B\'atiz esq. Av. Miguel Oth\'on de Mendiz\'abal, Col. Lindavista,
Delegaci\'on Gustavo A. Madero, C.P. 07738, Ciudad de M\'exico, Mexico.\\

\small $^{b}$ Escuela Superior de Ingenier{\'i}a Mec\'anica y El\'ectrica, Unidad Culhuac\'an,
Instituto Polit\'ecnico Nacional, Av. Santa Ana No. 1000, Col. San Francisco Culhuac\'an, Delegaci\'on Coyoac\'an, C.P. 04430, Ciudad de M\'exico, Mexico.\\

\small $^{c}$ Escuela Superior de F{\'i}sica y Matem\'aticas,
Instituto Polit\'ecnico Nacional, Ed. 9, Unidad Profesional Adolfo L\'opez Mateos, Delegaci\'on Gustavo A. Madero, C.P. 07738, Ciudad de M\'exico, Mexico.\\

\end{minipage}

(Date of resubmission: 1 June 2016.)

\begin{abstract}
We introduce a generalization of the Jaynes-Cummings model and study some of its properties. We obtain the energy
spectrum and eigenfunctions of this model by using the tilting transformation and the squeezed number states of the
one-dimensional harmonic oscillator. As physical applications, we connect this new model to two
important and novelty problems: the relativistic parametric amplifier and the quantum
simulation of a single trapped ion.

\end{abstract}


\section{Introduction}

The Dirac oscillator was first introduced by Ito \emph{et al.} \cite{Ito} and Cook \cite{Cook}
and reintroduced in the 80's by Moshinsky and Szczepaniak \cite{Mos}. They added the linear term
$-imc\omega\beta{\mathbf{\alpha}}\cdot \mathbf{r}$ to the relativistic momentum $\textbf{p}$ of the free-particle Dirac equation.
This problem reduces to the harmonic oscillator plus a spin-orbit coupling term in the non-relativistic limit.

The Dirac-Moshinsky oscillator has been applied to quark confinement models in quantum chromodynamics, hexagonal lattices
and the emulation of graphene in electromagnetic billiards \cite{Sadurni1}, among others. In particular,
in the $1+1$-dimensional space, the Dirac-Moshinsky oscillator has been exactly solved by using the theory of the non-relativistic harmonic
oscillator \cite{Nogami,Rados}. Moreover, Nogami and Toyama constructed the relativistic coherent state for this lower dimensional case \cite{Toyama}.
The $2+1$-dimensional Dirac-Moshinsky oscillator has been related to quantum optics via the Jaynes-Cummings and Anti-Jaynes-Cummings model \cite{Bermudez,Sadurni}.

In quantum optics the Jaynes-Cummings model describes the interaction between a two-level atom with a quantized electromagnetic field \cite{Jaynes}.
In the rotating wave approximation this model have been extensively studied and its exact solution has been found \cite{Haroche}. These solutions yield
quantum collapse and revival of atomic inversion \cite{Narozhny}, and squeezing of the radiation field \cite{Kuklinski}, among other quantum effects.
All these effects have been corroborated experimentally, as can be seen in references \cite{Goy,Brune,Guerlin}. In recent years, there have been published some works on simulations of Dirac equation using
different physical systems \cite{Lamata,Gerritsma,Lamata2}. These simulations allow to study relevant quantum relativistic effects,
like the Zitterbewegung and Klein's paradox. In reference  \cite{Johanning}, an extensive overview of
current theoretical proposals and experiments for such quantum simulations with trapped
ions is given.

The aim of the present work is to introduce a generalization of the Jaynes-Cummings model. We use pure
algebraic methods to obtain the energy spectrum and eigenfunctions. We show that this new model can
be reduced to the degenerate parametric amplifier in the non-relativistic limit. Also, our model is applied to the simulation of a single trapped ion.

This work is organized as it follows. In Section $2$, the main properties of the Dirac-Moshinsky oscillator in
$1+1$ dimensions and its mapping to the Anti-Jaynes-Cummings model are revisited. In Section $3$, we introduce the generalization of the Jaynes-Cummings model
in terms of two complex parameters and the creation and annihilation operators of the one-dimensional harmonic oscillator. We decouple the equations
for the upper and lower wave functions. By using the tilting transformation and the theory of the
one-dimensional harmonic oscillator we find its energy spectrum and eigenfunctions. In Section 4, for particular choices of the complex parameters, we apply our model to the
relativistic degenerate parametric amplifier and to a  single ion inside a Paul trap. Finally, we
give some concluding remarks.

\section{The $1+1$ Dirac-Moshinsky oscillator and the Anti-Jaynes-Cummings model.}

The time-independent Dirac equation for the Dirac-Moshinsky oscillator is given by the Hamiltonian \cite{Mos}
\begin{equation}
H_D\Psi=\left[c \mathbf{\alpha} \cdot\left(\mathbf{p}-im\omega \mathbf{r}\beta\right)+mc^2\beta\right]\Psi=E\Psi,
\end{equation}
where the Dirac matrices $\alpha$ and $\beta$ satisfy the Clifford algebra
\begin{align}\nonumber
\alpha_a\alpha_b+\alpha_b\alpha_a=2\delta_{ab}\textbf{1},\\
\alpha_a\beta+\beta\alpha_a=0,\\\nonumber
\alpha_a^2=\beta^2=\texttt{1}.
\end{align}
In $1+1$ dimensions, the most convenient representations of $\alpha$ and $\beta$ are
\begin{equation}
\alpha=\begin{pmatrix}
0 & -i \\
i & 0 \end{pmatrix},\quad\quad \beta=\begin{pmatrix}
1 & 0 \\
0 & -1 \end{pmatrix}.
\end{equation}
With this realization we obtain the coupled equations
\begin{equation}
\left(E-mc^2\right)|\Psi_1\rangle=c(-ip_x+m\omega x)|\Psi_2\rangle,\label{c1}
\end{equation}
\begin{equation}
\left(E+mc^2\right)|\Psi_2\rangle=c(ip_x+m\omega x)|\Psi_1\rangle.\label{c2}
\end{equation}
The uncoupled equations for the two components $|\Psi_1\rangle$ and $|\Psi_2\rangle$ can be
solved by using the non-relativistic theory of the quantum harmonic oscillator. The eigenvalues are \cite {Nogami,Rados}
\begin{equation}
E_n=\pm mc^2\sqrt{1+\frac{2|n|\hbar\omega}{mc^2}}, \quad\quad n=0,\pm1,\pm2,...\label{spectrum1}
\end{equation}
where the upper sign should be chosen for $n\geq0$ and the lower one for $n<0$. The normalized
eigenfunction $|\Psi\rangle$ is
\begin{equation}
|\Psi\rangle=\begin{pmatrix}
\sqrt{\frac{\lambda(E+mc^2)}{2^{|n|+1}|n|!\sqrt{\pi}E}}H_{|n|}(\lambda x)e^{-\lambda^2x^2/2}\\
\sqrt{\frac{\lambda(E-mc^2)}{2^{|n|}(|n|-1)!\sqrt{\pi}E}}H_{|n|-1}(\lambda x)e^{-\lambda^2x^2/2}
\end{pmatrix},
\end{equation}
where $H_n(x)$ is the Hermite polynomial and $\lambda=\sqrt{\frac{m\omega}{\hbar}}$.
If we introduce the usual creation and annihilation operators
\begin{equation}
a=\sqrt{\frac{m\omega}{2\hbar}}x+\frac{i}{\sqrt{2m\omega\hbar}}p_x,\quad\quad
a^{\dag}=\sqrt{\frac{m\omega}{2\hbar}}x-\frac{i}{\sqrt{2m\omega\hbar}}p_x,
\end{equation}
the coupled equations (\ref{c1}) and (\ref{c2}) can be written in the simplified form \cite{Toyama}
\begin{equation}
\left(E-mc^2\right)|\Psi_1\rangle=c\sqrt{2m\omega\hbar}a^{\dag}|\Psi_2\rangle,\label{c3}
\end{equation}
\begin{equation}
\left(E+mc^2\right)|\Psi_2\rangle=c\sqrt{2m\omega\hbar}a|\Psi_1\rangle.\label{c4}
\end{equation}
From these equations we can rewrite the Hamiltonian of the $1+1$-dimensional Dirac-Moshinsky oscillator as
\begin{equation}
H=\delta(\sigma_-a+\sigma_+a^{\dag})+mc^2\sigma_z,\label{MO}
\end{equation}
where $\delta=c\sqrt{2m\omega\hbar}$, $\sigma_+$ and $\sigma_-$ are the spin raising and lowering operators, and $\sigma_z$ is the
Pauli matrix. The equation (\ref{MO}) formally is the Hamiltonian of the Anti-Jaynes-Cummings model of the quantum optics.
Similarly, the $2+1$-dimensional Dirac-Moshinsky oscillator has been mapped onto the Anti-Jaynes-Cummings oscillator by
using the chiral creation and annihilation operators \cite{Bermudez}. The same conclusion was obtained
for the $2+1$-dimensional Dirac-Moshinsky oscillator coupled to an external magnetic field, by using the complex coordinate
$z$ and its conjugate momentum $p_z$ \cite{Mandal}.

\section{The generalized Jaynes-Cummings model.}

As we mentioned in the previous section, the Jaynes-Cummings
\begin{equation}
H=\hbar(g\sigma_-a^{\dag}+g^*\sigma_+a)+mc^2\sigma_z,\label{JC}
\end{equation}
and the Anti-Jaynes-Cummings
\begin{equation}
H=\hbar(g\sigma_-a+g^*\sigma_+a^{\dag})+mc^2\sigma_z,\label{AJC}
\end{equation}
models are very important in quantum optics. Therefore, we can propose a generalization of the Jaynes-Cummings model, in order to study and connect it with more general and complicated systems, besides the Dirac oscillator.
Thus, we introduce the generalized Jaynes-Cummings model as
\begin{equation}
H=\hbar\left[\sigma_-(g^*a+f a^{\dag})+\sigma_+(ga^{\dag}+f^*a)\right]+mc^2\sigma_z,\label{GAJC}
\end{equation}
where $g$ and $f$ are two general complex parameters. Notice that this model can be seen as a linear combination of the two Jaynes-Cummings models.

The coupled equations for the
spinor components $|\Psi_1\rangle$ and $|\Psi_2\rangle$ are
\begin{equation}
\hbar(ga^{\dag}+f^*a)|\Psi_2\rangle=(E-mc^2)|\Psi_1\rangle,\label{nc1}
\end{equation}
\begin{equation}
\hbar(g^*a+f a^{\dag})|\Psi_1\rangle=(E+mc^2)|\Psi_2\rangle.\label{nc2}
\end{equation}
These equations are the generalization of equations (\ref{c3}) and (\ref{c4}).
The uncoupled equations for $|\Psi_1\rangle$ and $|\Psi_2\rangle$ are easily obtained from above
expressions, which result to be
\begin{equation}
\hbar^2(|g|^2a^{\dag}a+|f|^2aa^{\dag}+gf a^{\dag^2}+f^*g^*a^2)|\Psi_1\rangle=(E^2-m^2c^4)|\Psi_1\rangle,\label{nu1}
\end{equation}
\begin{equation}
\hbar^2(|g|^2aa^{\dag}+|f|^2a^{\dag}a+gf a^{\dag^2}+f^*g^*a^2)|\Psi_2\rangle=(E^2-m^2c^4)|\Psi_2\rangle.\label{nu2}
\end{equation}
By using the $SU(1,1)$ Lie algebra realization in terms of the Bose operators $a$, $a^{\dag}$
\begin{equation}
K_0=\frac{1}{2}\left(a^{\dag}a+\frac{1}{2}\right), \quad K_+=\frac{1}{2}a^{\dag^2}, \quad K_-=\frac{1}{2}a^2,\label{gen}
\end{equation}
we can write the uncoupled equation for $|\Psi_1\rangle$ as
\begin{equation}
\left[2K_0\hbar^2(|g|^2+|f|^2)+2gf\hbar^2K_++2g^*f^*\hbar^2K_-+\frac{\hbar^2}{2}(|f|^2-|g|^2)\right]|\Psi_1\rangle=(E^2-m^2c^4)|\Psi_1\rangle\label{unc1},
\end{equation}
where we have used the property $aa^{\dag}=a^{\dag}a+1$. In order to remove the operators $K_{\pm}$, we apply the tilting transformation to the above Klein-Gordon-type Hamiltonian $H_{KG}|\Psi\rangle=\left(E^2-m^2 c^4\right)|\Psi\rangle$. To do this we proceed as in references \cite{gerryberry,Nos1}. Since
$D(\xi)D^{\dag}(\xi)=1$, equation (\ref{unc1}) can be written as
\begin{eqnarray}
D^{\dagger}(\xi)\left[2K_0\hbar^2(|g|^2+|f|^2)+2gf\hbar^2K_++2g^*f^*\hbar^2K_-+\frac{\hbar^2}{2}(|f|^2-|g|^2)\right]D(\xi)D^{\dagger}(\xi)|\Psi_1\rangle\nonumber\\=(E^2-m^2c^4)D^{\dagger}(\xi)|\Psi_1\rangle,
\end{eqnarray}
where $D(\xi)$ is the $SU(1,1)$ displacement or squeezing operator and $\xi=-\frac{1}{2}\tau e^{-i\varphi}$ (see Appendix). If we define the tilted Hamiltonian $H'_{KG}=D^{\dagger}(\xi)H_{KG}D(\xi)$ and the wave function $|\Psi'_1\rangle=D^{\dagger}(\xi)|\Psi_1\rangle$, this equation can be written as $H'_{KG}|\Psi'_1\rangle=(E^2-m^2c^4)|\Psi'_1\rangle$.

By choosing the coherent state parameter as $\tau=\tanh^{-1}\left(4|f||g|/(|f|^2+|g|^2)\right)$, and the definitions above, equation (\ref{unc1}) is transformed to
\begin{equation}
H'_{KG}|\Psi'_1\rangle=\hbar^2\left(2K_0-\frac{1}{2}\right)\left(|g|^2-|f|^2\right)|\Psi'_1\rangle=(E^2-m^2c^4)|\Psi'_1\rangle,\label{hkg}
\end{equation}
where $2K_0$ is the Hamiltonian of the one-dimensional harmonic oscillator. Since the
energy spectrum of the one-dimensional harmonic oscillator is $n+\frac{1}{2}$, we obtain the energy spectrum of the generalized Jaynes-Cummings model
\begin{equation}
E=\pm\sqrt{\hbar^2(|g|^2-|f|^2)n+m^2c^4}.\label{spectrum}
\end{equation}
If we apply the same procedure to the uncoupled equation for the other spinor component $|\Psi_2\rangle$ we obtain the energy spectrum
\begin{equation}
E=\pm\sqrt{\hbar^2(|g|^2-|f|^2)(n'+1)+m^2c^4}.
\end{equation}
Thus, since both components $|\Psi_1\rangle$ and $|\Psi_2\rangle$ belong the same energy, we obtain that the relationship between
the quantum numbers is $n=n'+1$. Moreover, if we set $f=0$ in the generalized Jaynes-Cummings model of equation (\ref{GAJC}),
the standard Dirac-Moshinsky oscillator is recovered. Also, under this setting, if we identify $\delta=g\hbar$, the spectrum (\ref{spectrum})
is simplified to that of equation (\ref{spectrum1}).

The eigenfunctions of the tilted Hamiltonian $H'_{KG}$ are those of the one-dimensional harmonic oscillator
\begin{equation}
\Psi'_1(x)=\sqrt{\frac{1}{\pi^{1/4}(2^n n!)^{1/2}}}e^{-\frac{1}{2}x^2}H_n(x),
\end{equation}
where $H_n(x)$ are the Hermite polynomials. Therefore, the eigenfunctions of the generalized Jaynes-Cummings model are obtained from $|\Psi_1\rangle=D(\xi)|\Psi'_1\rangle$.  Similar results hold for $|\Psi_1\rangle$.  These states are known as the squeezed number states of the one-dimensional harmonic oscillator \cite{Nieto2}. The action of the squeezing operator $D(\xi)$ on $|\Psi'_1\rangle$  is a long calculation. However it has been calculated by Nieto  in reference  \cite{Nieto2} (see Appendix).
 By using these  results we are able to construct  the spinor $|\Psi\rangle$ in the form
\begin{equation}
|\Psi\rangle=\frac{1}{\pi^{1/4}(F_1)^{1/2}}e^{\left(-\frac{1}{2}x^2F_2\right)}\begin{pmatrix}
A_n\frac{(F_3)^{n/2}}{(2^nn!)^{1/2}}H_n\left(\frac{x}{F_4}\right)\\
B_n\frac{(F_3)^{(n-1)/2}}{(2^{n-1}(n-1)!)^{1/2}}H_{n-1}\left(\frac{x}{F_4}\right)
\end{pmatrix},\quad n=1,2,...
\end{equation}
where $F_1$,  $F_2$, $ F_3$ and $F_4$ are functions depending on the coherent state parameters, given in the Appendix.
Since the squeezed number states are already normalized, after normalization we obtain that the spinor $|\Psi\rangle$ is
\begin{equation}
|\Psi\rangle=\frac{1}{\pi^{1/4}(F_1)^{1/2}}e^{\left(-\frac{1}{2}x^2F_2\right)}\begin{pmatrix}
\sqrt{\frac{E\pm mc^2}{2E}}\frac{(F_3)^{n/2}}{(2^nn!)^{1/2}}H_n\left(\frac{x}{F_4}\right)\\
\mp i\sqrt{\frac{E\mp mc^2}{2E}}\frac{(F_3)^{(n-1)/2}}{(2^{n-1}(n-1)!)^{1/2}}H_{n-1}\left(\frac{x}{F_4}\right)
\end{pmatrix}.\label{estado}
\end{equation}

Therefore, we have solved the generalized Jaynes-Cummings model, which includes the Jaynes-Cummings and Anti-Jaynes-Cummings interactions, by using
the tilting transformation and the squeezed states of the one-dimensional harmonic oscillator.

\section{Special cases: The relativistic parametric amplifier and a single trapped ion.}

In this section, we shall give two physical applications of the theory developed in the last section related to the generalized Jaynes-Cummings model. First,
we shall give the connection between the generalized Jaynes-Cummings model and the relativistic parametric amplifier. Next, we shall apply our model to a single ion inside a Paul trap.

\subsection{The relativistic degenerate parametric amplifier}

A particular case of the Hamiltonian proposed in equation (\ref{GAJC}), is obtained if we
set the complex parameters $g$ and $f$ as
\begin{equation}
g=\frac{imc^2\sqrt{2\mu}}{\hbar}e^{-i\phi}, \quad\quad f=\frac{\chi}{\sqrt{2\mu}}e^{-i\phi}, \quad\quad \mu=\frac{\hbar\omega}{mc^2}.\label{part}
\end{equation}
The physical meaning of $\chi$ and the phase $\phi$ will be clarified below on this section. Thus, the uncoupled equation for the upper component $|\Psi_1\rangle$ takes the form (see equation (\ref{nu1}))
\begin{equation}
H_{KG}|\Psi_1\rangle=(E^2-m^2c^4)|\Psi_1\rangle, \label{pnu1}
\end{equation}
where the Klein-Gordon-type Hamiltonian written in terms of the one-dimensional $SU(1,1)$  harmonic oscillator generators of equation (\ref{gen}) is
\begin{equation}
H_{KG}=\hbar^2\left[\left(\frac{2mc^2\omega}{\hbar}+\frac{mc^2\chi^2}{2\hbar\omega}\right)a^{\dag}a+\frac{mc^2\chi^2}{2\hbar\omega}
-\frac{imc^2\chi}{\hbar}\left(a^2e^{2i\phi}-a^{\dag^2}e^{-2i\phi}\right)\right].
\end{equation}
A similar equation holds for the lower component $|\Psi_2\rangle$. The transformations performed to the general
case (equations (\ref{unc1})-(\ref{hkg})), diagonalize this Hamiltonian to
\begin{equation}
H'_{KG}|\Psi'_1\rangle=\hbar^2\left(2a^{\dag}a+\frac{1}{2}\right)\left(\frac{2mc^2\omega}{\hbar}-\frac{mc^2\chi^2}{2\hbar\omega} \right)|\Psi'_1\rangle=(E^2-m^2c^4)|\Psi'_1\rangle.
\end{equation}
The exact energy spectrum for $g$ and $f$ given by (\ref{part}) is obtained from equation (\ref{spectrum})
\begin{equation}
E=\pm mc^2\sqrt{1+\left(\frac{2\hbar\omega}{mc^2}-\frac{\hbar\chi^2}{2\omega mc^2}\right)n},\label{jcspec}
\end{equation}
and the corresponding eigenfunctions are given by equation (\ref{estado}).

By written $E=mc^2+\epsilon$, we obtain that the non-relativistic limit $(\epsilon\ll mc^2)$ of the uncoupled equation (\ref{pnu1}) becomes
\begin{equation}
\left[\left(\hbar\omega+\frac{\hbar\chi^2}{4\omega}\right)a^{\dag}a+\frac{\hbar\chi^2}{4\omega}-i\hbar\frac{\chi}{2}\left(a^2e^{2i\phi}
-a^{\dag^2}e^{-2i\phi}\right)\right]|\Psi_1\rangle=\epsilon|\Psi_1\rangle.
\end{equation}
Formally, the operator on the left hand is the Hamiltonian of the time-independent degenerate parametric amplifier \cite{gerryberry,walls}, which  has been  used as a model for producing non-classical states of the quantized electromagnetic field. It is at this level where we can give the  physical interpretation of the parameters $\chi$ and $\phi$. $\chi$ is the coupling constant (proportional to the second-order susceptibility of the medium and to the amplitude of the pump), $\phi$ is the phase of the pump field and $\omega$ is its frequency \cite{gerryberry,walls}. Moreover, the relativistic energy spectrum, equation (\ref{jcspec}), in the non-relativistic limit is in full agreement with the energy spectrum of the degenerate parametric amplifier \cite{gerryberry}. Also,  if we set $\chi=0$ in equation (\ref{jcspec}), the energy spectrum is reduced to that of the $1+1$-dimensional Dirac-Moshinsky oscillator of equation (\ref{spectrum1}). Therefore, the generalized Jaynes-Cummins model (equation (\ref{GAJC})) is a relativistic version of the degenerate parametric amplifier.

It is important to note that the parameters $f$ and $g$ of equation (\ref{part}) can be also defined as
\begin{equation}
g=\frac{imc^2\sqrt{2\mu}}{\hbar}, \quad\quad f=\frac{\chi}{\sqrt{2\mu}}e^{-2i\phi}, \quad\quad \mu=\frac{\hbar\omega}{mc^2}.
\end{equation}
With this definition $g$ is now purely imaginary. It can be easily shown that these parameters lead to the same relativistic version (equations (\ref{pnu1})-(\ref{jcspec})) of the degenerate parametric amplifier.

In references \cite{Bermudez,Mandal} it has been shown that the $2+1$ Dirac oscillator can be exactly mapped to an Jaynes or Anti-Jaynes-Cummings interaction. In the present work
we showed that the generalized Jaynes-Cummings model, which includes the Jaynes and Anti-Jaynes-Cummings interaction, can be mapped to the degenerate parametric amplifier in the
non-relativistic limit.

\subsection{Quantum simulation of a single ion inside a Paul trap}

In references \cite{Bermudez,Lamata} the authors proposed an experiment in order
to study the equivalence between the Dirac oscillator and the Jaynes-Cummings model.
This experiment corresponds to a single ion of mass $M$ inside a Paul trap. In this context, the Jaynes-Cummings (JC) interaction is known as the red-sideband excitation, and consists
of a laser field acting resonantly on two internal levels and one vibrational mode. The Anti-Jaynes-Cummings (AJC) interaction consists of a JC-like coupling tuned
to the blue motional sideband \cite{Lamata}. A carrier interaction coherently couples two ionic internal levels without changing the external motion
of the ion. The corresponding interaction Hamiltonians under the rotating wave and Lamb-Dicke approximations are
\begin{equation}
H_r=\hbar \eta\tilde{\Omega}\left(\sigma_+ae^{i\phi_r}+\sigma_-a^{\dag}e^{-i\phi_r}\right), \quad\quad H_b=\hbar \eta\tilde{\Omega}\left(\sigma_+a^{\dag}e^{i\phi_b}+\sigma_-ae^{-i\phi_b}\right),\nonumber
\end{equation}
\begin{equation}
H_{\sigma}=\hbar \eta\Omega\left(\sigma_+e^{i\phi}+\sigma_-e^{-i\phi}\right),
\end{equation}
where $\tilde{\Omega}$ is the coupling strength ($\Omega$ for the carrier transition), $\eta=\sqrt{\frac{\hbar k^2}{2M\nu}}$ is the Lamb-Dicke parameter and $k$ is the wave number of the driving field.
When the blue and red sideband couplings are used simultaneously, the resulting Hamiltonian is the sum of $H_r$ and $H_b$ \cite{Johanning}
\begin{equation}
H=\hbar \eta\tilde{\Omega}\left[\sigma_+\left(ae^{i\phi_r}+a^{\dag}e^{i\phi_b}\right)+\sigma_-\left(a^{\dag}e^{-i\phi_r}+ae^{-i\phi_b}\right)\right].\label{ion}
\end{equation}
If we now set the complex parameters $g$ and $f$ as
\begin{equation}
g=\eta \tilde{\Omega}e^{i\phi_b}, \quad\quad f=\eta \tilde{\Omega}e^{i\phi_r},\label{part2}
\end{equation}
the generalized Jaynes-Cummings model, equation (\ref{GAJC}), turns out to be the Hamiltonian of an ion inside a Paul trap of equation (\ref{ion}).
Therefore, by substituting these complex parameters into equation (\ref{spectrum}), we obtain that the energy of the ion trapped
is just the relativistic rest mass energy $E=\pm mc^2$. This important result is ought to the Doppler cooling.

The importance of this model lies in the fact that in the experiment proposed in references \cite{Bermudez,Lamata} some relevant quantum-relativistic
effects appear, like the \emph{Zitterbewegung}, the Klein's paradox and collapse-revival dynamics. In reference \cite{Gerritsma}, the authors
simulate a single trapped $^{40}Ca^+$ ion in a linear Paul trap and investigate the particle dynamics in the crossover from relativistic
to non-relativistic dynamics.

\section{Concluding remarks}

We introduced a generalization of the Jaynes-Cummings model in terms of the creation and annihilation operators
of the one-dimensional harmonic oscillator. This model is a linear combination of the Jaynes-Cummings and Anti-Jaynes-Cummings models.
We obtained the exact solution of this problem by using the tilting transformation and the squeezed number states of the
one-dimensional harmonic oscillator.

For a particular choice of our model parameters, we constructed the relativistic version of the degenerate parametric amplifier. We obtained the exact energy spectrum of this relativistic model
and we showed that when the coupling constant $\chi=0$, the energy spectrum of the $1+1$-dimensional Dirac-Moshinsky oscillator is recovered. Also, with a different choice of the parameters, we connected
our generalized Jaynes-Cummings model with the Hamiltonian of the quantum simulation of a single trapped ion.

The Jaynes-Cummings model has been studied by using a different algebraic approach in reference \cite{Rauch}. In this reference,
the authors introduced a Jaynes-Cummings model which also describes spin-orbit interaction and used a representation of the $su(2)$
algebra which arises naturally from the system. This matrix $su(2)$ Lie algebra is constructed by a properly combination of the operators
of a generalized deformed oscillator algebra $\{A_0,A_+,A_-$\}, and the well-known Pauli matrices $\sigma_x,\sigma_y,\sigma_z$. Based on
these ideas, it can be proposed an analogous approach to find the algebraic structure of the Anti-Jaynes-Cummings model. In fact, it can be
shown that a different combination of the operators $\{A_0,A_+,A_-$\}, and the Pauli matrices $\sigma_x,\sigma_y,\sigma_z$ allows to produce
an $su(2)$ symmetry for the Anti-Jaynes-Cummins Hamiltonian. Thus, the Jaynes-Cummings and Anti-Jaynes-Cummings possess an $su(2)$ hidden
algebraic structure. These results allow to think that the generalized Jaynes-Cummings could be possesses an $su(2)$ symmetry also.

The generalization of the Jaynes-Cummings model introduced in this work was constructed with only one oscillator.
The formulation developed in this work can be extended to a model that includes two oscillators. This new generalization
shall permit us to have a wider range of applications and will be reported in a future work.

\section*{Acknowledgments}
The authors thank the referee's suggestions to improve our paper. This work was partially supported by SNI-M\'exico, COFAA-IPN,
EDI-IPN, EDD-IPN, SIP-IPN project number $20161727$.

\section{Appendix. $su(1,1)$ Lie algebra and its squeezed and number coherent states}

Three operators $K_{\pm}, K_0$ close the $su(1,1)$ Lie algebra if they satisfy the commutation relations \cite{Vourdas}
\begin{eqnarray}
[K_{0},K_{\pm}]=\pm K_{\pm},\quad\quad [K_{-},K_{+}]=2K_{0}.\label{com}
\end{eqnarray}
The action of these operators on the Fock space states
$\{|k,n\rangle, n=0,1,2,...\}$ is
\begin{equation}
K_{+}|k,n\rangle=\sqrt{(n+1)(2k+n)}|k,n+1\rangle,\label{k+n}
\end{equation}
\begin{equation}
K_{-}|k,n\rangle=\sqrt{n(2k+n-1)}|k,n-1\rangle,\label{k-n}
\end{equation}
\begin{equation}
K_{0}|k,n\rangle=(k+n)|k,n\rangle,\label{k0n}
\end{equation}
where $|k,0\rangle$ is the lowest normalized state. The Casimir
operator $K^{2}$ for any irreducible representation of this group is given by
\begin{equation}
K^2=K^2_0-\frac{1}{2}(K_+K_-+K_-K_+)
\end{equation}
and satisfies the relationship $K^{2}=k(k-1)$. Thus, a representation of
$su(1,1)$ algebra is determined by the number $k$, called the Bargmann index. The discrete series
are those for which $k>0$.

The $SU(1,1)$ Perelomov coherent states are defined as the action of the displacement operator $D(\xi)$
onto the lowest normalized state $|k,0\rangle$ as \cite{Perellibro}
\begin{equation}
|\zeta\rangle=D(\xi)|k,0\rangle=(1-|\zeta|^2)^k\sum_{n=0}^\infty\sqrt{\frac{\Gamma(n+2k)}{n!\Gamma(2k)}}\zeta^n|k,n\rangle,\label{PCN}
\end{equation}
The displacement operator $D(\xi)$ is defined in terms of the creation and annihilation operators $K_+, K_-$ as
\begin{equation}
D(\xi)=\exp(\xi K_{+}-\xi^{*}K_{-}),\label{do}
\end{equation}
where $\xi=-\frac{1}{2}\tau e^{-i\varphi}$, $-\infty<\tau<\infty$ and $0\leq\varphi\leq2\pi$.
The so-called normal form of the squeezing operator is given by
\begin{equation}
D(\xi)=\exp(\zeta K_{+})\exp(\eta K_{0})\exp(-\zeta^*K_{-})\label{normal},
\end{equation}
where  $\zeta=-\tanh(\frac{1}{2}\tau)e^{-i\varphi}$ and $\eta=-2\ln \cosh|\xi|=\ln(1-|\zeta|^2)$ \cite{Gerry}.

The $SU(1,1)$ Perelomov number coherent state $|\zeta,k,n\rangle$ is defined as the action of the displacement operator $D(\xi)$ onto an arbitrary
excited state $|k,n\rangle$ \cite{Nos1}
\begin{eqnarray}
|\zeta,k,n\rangle &=&\sum_{s=0}^\infty\frac{\zeta^s}{s!}\sum_{j=0}^n\frac{(-\zeta^*)^j}{j!}e^{\eta(k+n-j)}
\frac{\sqrt{\Gamma(2k+n)\Gamma(2k+n-j+s)}}{\Gamma(2k+n-j)}\nonumber\\
&&\times\frac{\sqrt{\Gamma(n+1)\Gamma(n-j+s+1)}}{\Gamma(n-j+1)}|k,n-j+s\rangle.\label{PNCS}
\end{eqnarray}

It can be seen that the operators
\begin{equation}
K_0=\frac{1}{2}\left(a^{\dag}a+\frac{1}{2}\right), \quad K_+=\frac{1}{2}a^{\dag^2}, \quad K_-=\frac{1}{2}a^2.\label{opho}
\end{equation}
satisfy the $su(1,1)$ Lie algebra commutation relations (\ref{com}). In this expressions $a$ and $a^{\dag}$ are the creation
and annihilation operators of the one-dimensional harmonic oscillator. The number states of the harmonic oscillator decompose into
two invariant subspaces, spanned by even number states $|2n\rangle$ and odd numbers states $|2n+1\rangle$, $n=0,1,...$.
For this case, the displacement operator $D(\xi)$ is called squeeze operator $S(\xi)$ and is given by
\begin{equation}
D(\xi)=S(z)=e^{\left(\frac{1}{2}z a^{\dag^2}-\frac{1}{2}z^*a^2\right)}, \quad\quad z=re^{i\varphi}.
\end{equation}
If we apply the squeeze operator to a number state $|n\rangle$ (for n even) of the harmonic oscillator we obtain
\begin{equation}
S(z)|n_e\rangle=\frac{1}{\pi^{1/4}(F_1)^{1/2}}e^{\left(-\frac{1}{2}x^2F_2\right)}\left[\frac{(F_3)^{n/2}}{(2^nn!)^{1/2}}H_n\left(\frac{x}{F_4}\right)\right],\label{sns}
\end{equation}
where $H_n$ are the Hermite polynomials and $F_1$, $F_2$, $F_3$ and $F_4$ are functions of the coherent state parameters. These functions and are given explicitly by \cite{Nieto2}
\begin{equation}
F_1=\cosh{r}+e^{i\varphi}\sinh{r},
\end{equation}
\begin{equation}
F_2=\frac{1-i\sinh{\varphi}\sinh{r}(\cosh{r}+e^{i\varphi}\sinh{r})}{(\cosh{r}+\cos{\varphi}\sinh{r})(\cosh{r}+e^{i\varphi}\sinh{r})},
\end{equation}
\begin{equation}
F_3=\frac{\cosh{r}+e^{-i\varphi}\sin{\varphi}\sinh{r}}{\cosh{r}+e^{i\varphi}\sin{\varphi}\sinh{r}},
\end{equation}
\begin{equation}
F_4=\left(\cosh^2{r}+\sinh^2{r}+2\cos{\varphi}\cosh{r}\sinh{r}\right)^{1/2}.
\end{equation}
A similar expression is valid when $n$ is odd.

\end{document}